\documentclass[showpacs,preprintnumbers]{revtex4}
\usepackage{graphicx}
\usepackage{color}

\def\bc{\begin{center}}
\def\ec{\end{center}}

\def\beq{\begin{equation}}
\def\eeq{\end{equation}}

\begin{document}

\title{Quantum entanglement  for two qubits in a nonstationary cavity}

\author{Oleg L. Berman$^{1,2}$, Roman Ya. Kezerashvili$^{1,2}$,  and Yurii E.
Lozovik$^{3,4}$}
 \affiliation{\mbox{$^{1}$Physics Department, New York
City College
of Technology, The City University of New York,} \\
Brooklyn, NY 11201, USA \\
\mbox{$^{2}$The Graduate School and University Center, The
City University of New York,} \\
New York, NY 10016, USA \\
\mbox{$^{3}$Institute of Spectroscopy, Russian Academy of Sciences,
142190 Troitsk, Moscow, Russia }\\
\mbox{$^{4}$MIEM at National Research University Higher School of
Economics, Moscow, Russia}}

\date{\today}

\begin{abstract}
 The quantum entanglement and the probability of the dynamical
Lamb effect  for two  qubits   caused by non-adiabatic fast change of the boundary conditions are
 studied. The conditional concurrence of the qubits for each fixed number of created photons  in a nonstationary cavity
is obtained as  a measure of the dynamical quantum entanglement due to the dynamical Lamb effect.  We discuss the physical
realization of the dynamical Lamb effect, based on superconducting qubits.
\end{abstract}


\maketitle

\section{Introduction}

\label{intro}

The new direction in the quantum electrodynamics (QED), related to
the optical properties of the cavity with the nonstationary
boundary conditions, initiated studies of several interesting
physical phenomena such as a dynamical Casimir effect
(DCE)~\cite{Moore,Fulling,Dodonov_pscr,Dodonov_pra}.    Casimir
predicted that two mirrors, which are perfectly conducting metal
plates, held parallel to each other in vacuum, will experience an
attractive force~\cite{Casimir}. Essentially, the mirrors reduce the
density of the electromagnetic modes between them. The vacuum radiation
pressure between the plates is then less than the pressure outside,
that generates the attractive force. This effect of quantum electrodynamics is
known as the static Casimir effect. If the mirrors move sufficiently fast,  a mismatch between vacuum modes at different times appears. It has been
predicted that this results in creation of real photons out of
vacuum fluctuations called the dynamical Casimir effect (DCE)~\cite{Moore}.  The control  of
the Lamb shift  in a nonstationary cavity due to adiabatical change of the cavity modes was also considered~\cite{BLP}.
 The new effect of cavity quantum electrodynamics, the dynamical Lamb effect  (DLE), which is photonless, parametric excitation of an atom, embedded in a nonstationary
 cavity, by shaking its photonic coat due to nonadiabatic change of the boundary conditions for virtual
 photons, contributing to the electronic self-energy, was predicted~\cite{Lozovik}.

  The fast changes of the boundary
conditions can be  achieved in superconducting circuits. The
superconducting qubits play the role of  atoms in an optical
cavity, and the superconducting line plays the role of an optical
cavity~\cite{Girvin,Ustinov,Girvin_Nature,Nori_nphys,Nigg,Leghas}.
The fast changes of the boundary conditions can be achieved by using
superconducting quantum interference device (SQUID) with the fast
changes of the external magnetic field. It was shown that a coplanar
waveguide (CPW) terminated by a SQUID is a very promising system for
experimental observation of the dynamical Casimir effect  for the
$\mathrm{GHz}$ frequency range instead of optical region for
regular atoms in an optical cavity~\cite{Johansson,Wilson}.
  The observation of the DCE requires to go sufficiently beyond
the regime of  adiabatic change of the boundary conditions of the
cavity. Non-adiabatic change of these boundary conditions
corresponds to the requirement that the characteristic time of the
change of the boundary conditions must be smaller than the period of
oscillations of the electromagnetic field, formed due to the DCE.
This nonadiabatic change of the boundary conditions requires that
the frequency of the photon, excited due to the DCE, must be smaller
than the frequency of the change of the boundary conditions. The
high quality factor of the superconducting cavity and the
possibilities for easy control (using SQUID) of the boundary conditions with the
frequencies of the order of magnitude of the eigenfrequencies  of
the superconducting cavity  allow  to observe DCE and DLE
for the superconducting qubits.

In this paper, we consider the physical realization to observe the quantum entanglement and the DLE
for the artificial atoms, formed by
  superconducting qubits, constructed in a superconducting circuit based on Josephson
 junctions. We analyze  superconducting qubits,
 coupled to a CPW terminated by a
SQUID. Changing the magnetic flux through the SQUID parametrically
modulates the boundary conditions of CPW. Ultrafast change of CPW boundary conditions  results in the
excitation of the superconducting qubit which is identical to the dynamical Lamb effect for an ordinary atom
  in a cavity with the nonstationary changing parameters. We focus on the physical realization of the
DLE~\cite{Lozovik} resulting in excitation of two qubits.
While the influence of the modulation of the constant of the
qubit-photon interaction on  excitation of the qubit was studied
in Ref.~\onlinecite{Lozovik_2015},  below we propose
excitation of the qubits due to the change of the boundary conditions
of the cavity.   We analyze  the quantum
entanglement and obtain the probability of the DLE for two  qubits, coupled to an optical cavity.

The paper is organized in the following way. In Sec.~\ref{DLE}, we study the dynamical Lamb effect for
 two qubits, coupled to an optical cavity with changing boundary conditions, and derive the probability of the DLE.
 In Sec.~\ref{entan}, we analyze the quantum entanglement for two qubits
 caused by nonadiabatic fast change of the boundary conditions, and obtain  the conditional concurrence for two qubits at fixed number of
created photons in a nonstationary cavity. Finally, the discussion of the results and conclusions
follow in Sec.~\ref{disc}.

\section{The dynamical Lamb effect for two qubits, coupled to an optical cavity}

\label{DLE}

We consider two qubits, coupled to an optical cavity with changing boundary conditions.
The Hamiltonian $\hat{\mathcal{H}}$ of two qubits in a nonstationary cavity,   assuming $\hbar = 1$,  is
given by (the Hamiltonian of one qubit in a nonstationary cavity was presented in Ref.~\onlinecite{Lozovik})
%%%%%%%%%%%%%%%%%%%%%%%%%%%%%%%%%%%%%%%%%%%%%%%%%%%%%%%%%%%%%%%%%%%%%%%%%%%%%%%%%%%%%%%%%%%%%%%%%%%%%
\begin{eqnarray}
\label{ham22} \hat{\mathcal{H}} = \hat{H}  +  i \frac{\dot{\omega}(t)}{4 \omega(t)}\left(a^{2} - a^{\dagger 2}\right)\ ,
\end{eqnarray}
%%%%%%%%%%%%%%%%%%%%%%%%%%%%%%%%%%%%%%%%%%%%%%%%%%%%%%%%%%%%%%%%%%%%%%%%%%%%%%%%%%%%%%%%%%%%%%%%%%%%%
where $\hat{H}$ is the Hamiltonian of a stationary cavity, defined as
%%%%%%%%%%%%%%%%%%%%%%%%%%%%%%%%%%%%%%%%%%%%%%%%%%%%%%%%%%%%%%%%%%%%%%%%%%%%%%%%%%%%%%%%%%%%%%%%%%%%%
\begin{eqnarray}
\label{ham223}
\hat{H} = \hat{H}_{0} + \hat{V}_{total} \ ,
\end{eqnarray}
%%%%%%%%%%%%%%%%%%%%%%%%%%%%%%%%%%%%%%%%%%%%%%%%%%%%%%%%%%%%%%%%%%%%%%%%%%%%%%%%%%%%%%%%%%%%%%%%%%%%%
where $\hat{H}_{0}$ is the Hamiltonian of two qubits without
qubit-photon interaction, which is given by
%%%%%%%%%%%%%%%%%%%%%%%%%%%%%%%%%%%%%%%%%%%%%%%%%%%%%%%%%%%%%%%%%%%%%%%%%%%%%%%%%%%%%%%%%%%%%%%%%%%%%
\begin{eqnarray}
\label{ham022} \hat{H}_{0} =
E_{0}\sum_{j=1,2}\frac{1+\hat{\sigma}_{3j}}{2} + \omega a^{\dagger}a
\ ,
\end{eqnarray}
%%%%%%%%%%%%%%%%%%%%%%%%%%%%%%%%%%%%%%%%%%%%%%%%%%%%%%%%%%%%%%%%%%%%%%%%%%%%%%%%%%%%%%%%%%%%%%%%%%%%%
$\hat{V}_{total}$ is the total Hamiltonian of the qubit-photon interaction given
by
%%%%%%%%%%%%%%%%%%%%%%%%%%%%%%%%%%%%%%%%%%%%%%%%%%%%%%%%%%%%%%%%%%%%%%%%%%%%%%%%%%%%%%%%%%%%%%%%%%%%%
\begin{eqnarray}
\label{V22} \hat{V}_{total} = \lambda \sum_{j=1,2}
\left(\hat{\sigma}_{j}^{+} + \hat{\sigma}_{j}^{-}\right)\left(a +
a^{\dagger}\right) \ .
\end{eqnarray}
%%%%%%%%%%%%%%%%%%%%%%%%%%%%%%%%%%%%%%%%%%%%%%%%%%%%%%%%%%%%%%%%%%%%%%%%%%%%%%%%%%%%%%%%%%%%%%%%%%%%%
In Eqs.~(\ref{ham022}) and~(\ref{V22}), $\hat{\sigma}^{+} =
\left(\hat{\sigma}_{1} + i \hat{\sigma}_{2}\right)/2$,
$\hat{\sigma}^{-} = \left(\hat{\sigma}_{1} - i
\hat{\sigma}_{2}\right)/2$, assuming $\hat{\sigma}_{1}$,
$\hat{\sigma}_{2}$, and $\hat{\sigma}_{3}$ are the Pauli matrices,
$a^{\dagger}$ and $a$ are creation and annihilation operators for
cavity photons, correspondingly, $E_{0}$ is the qubit transition
frequency, $\lambda$ is the strength of the artificial qubit-photon
coupling. The second term in the r.h.s. of Eq.~(\ref{ham22}) was obtained in Ref.~\onlinecite{Law}.

It follows from Eq.~(\ref{V22}) that
%%%%%%%%%%%%%%%%%%%%%%%%%%%%%%%%%%%%%%%%%%%%%%%%%%%%%%%%%%%%%%%%%%%%%%%%%%%%%%%%%%%%%%%%%%%%%%%%%%%%%
\begin{eqnarray}
\label{V22totrwa} \hat{V}_{total} = \hat{V}_{RWA} +  \hat{V} \ ,
\end{eqnarray}
%%%%%%%%%%%%%%%%%%%%%%%%%%%%%%%%%%%%%%%%%%%%%%%%%%%%%%%%%%%%%%%%%%%%%%%%%%%%%%%%%%%%%%%%%%%%%%%%%%%%%
where $\hat{V}_{RWA}$ is the qubit-photon interaction in the rotating wave approximation (RWA)
%%%%%%%%%%%%%%%%%%%%%%%%%%%%%%%%%%%%%%%%%%%%%%%%%%%%%%%%%%%%%%%%%%%%%%%%%%%%%%%%%%%%%%%%%%%%%%%%%%%%%
\begin{eqnarray}
\label{V22rwa} \hat{V}_{RWA} = \lambda \sum_{j=1,2}
\left(\hat{\sigma}_{j}^{+}a + \hat{\sigma}_{j}^{-}a^{\dagger}\right) \ ,
\end{eqnarray}
%%%%%%%%%%%%%%%%%%%%%%%%%%%%%%%%%%%%%%%%%%%%%%%%%%%%%%%%%%%%%%%%%%%%%%%%%%%%%%%%%%%%%%%%%%%%%%%%%%%%%
which does not change the number of the excitations in the system, and  $\hat{V}$ is the term beyond the RWA, which can increase or decrease the number of
 the excitations in the system for two or four excitations, including the qubit and photon excitations:
 %%%%%%%%%%%%%%%%%%%%%%%%%%%%%%%%%%%%%%%%%%%%%%%%%%%%%%%%%%%%%%%%%%%%%%%%%%%%%%%%%%%%%%%%%%%%%%%%%%%%%
\begin{eqnarray}
\label{V22non} \hat{V} = \lambda \sum_{j=1,2}
\left(\hat{\sigma}_{j}^{+}a^{\dagger} + \hat{\sigma}_{j}^{-}a\right) \ .
\end{eqnarray}
%%%%%%%%%%%%%%%%%%%%%%%%%%%%%%%%%%%%%%%%%%%%%%%%%%%%%%%%%%%%%%%%%%%%%%%%%%%%%%%%%%%%%%%%%%%%%%%%%%%%%
Since the DLE is related to the parametric excitation of the qubit, only $\hat{V}$ influences the DLE.

The second term in Eq.~(\ref{ham22}) describes the dynamical Casimir effect, parametric excitation of photons due to non-adiabatical change of the cavity photon frequency.
There are three channels of qubit excitations in a nonstationary optical cavity due to non-adiabatical change of the cavity eigenfrequency. The first and second channels
correspond to excitation of a qubit by real and virtual Casimir photons, correspondingly. This paper is focused on the third channel,
 which corresponds to a new effect of nonstationary
quantum electrodynamics, the dynamical Lamb effect, which is parametric excitation of an atom due to  modulation of its Lamb shift or, in other words, due to  shaking
of its photonic coat.

Using  Eqs.~(\ref{ham022}) and~(\ref{V22}), one gets the eigenstate vectors of the Hamiltonian
$\hat{H}$ in the framework of the first order perturbation theory~\cite{Landau}:
%%%%%%%%%%%%%%%%%%%%%%%%%%%%%%%%%%%%%%%%%%%%%%%%%%%%%%%%%%%%%%%%%%%%%%%%%%%%%%%%%%%%%%%%%%%%%%%%%%%%%
\begin{eqnarray}
\label{st1} \left|n;0 0\right\rangle_{\lambda \omega} &=& \left|n;0
0\right\rangle + \frac{\lambda\sqrt{n}\left(\left|n-1;1
0\right\rangle + \left|n-1;0 1\right\rangle\right)}{\omega - E_{0}}
- \frac{\lambda\sqrt{n+1}\left(\left|n+1;1 0\right\rangle +
\left|n+1;0 1\right\rangle\right)}{\omega + E_{0}} \ , \nonumber \\
\left|n;1 0\right\rangle_{\lambda \omega} &=& \left|n;1
0\right\rangle + \frac{\lambda\sqrt{n}\left|n-1;0 0\right\rangle -
\lambda\sqrt{n+1}\left|n+1;1 1\right\rangle}{\omega + E_{0}} +
\frac{\lambda\sqrt{n}\left|n-1;1 1\right\rangle -
\lambda\sqrt{n+1}\left|n+1;0 0\right\rangle}{\omega - E_{0}}   \ , \nonumber \\
\left|n;0 1\right\rangle_{\lambda \omega} &=& \left|n;0
1\right\rangle + \frac{\lambda\sqrt{n}\left|n-1;0 0\right\rangle -
\lambda\sqrt{n+1}\left|n+1;1 1\right\rangle}{\omega + E_{0}} +
\frac{\lambda\sqrt{n}\left|n-1;1 1\right\rangle -
\lambda\sqrt{n+1}\left|n+1;0 0\right\rangle}{\omega - E_{0}}
\ , \nonumber \\
\left|n;1 1\right\rangle_{\lambda \omega} &=& \left|n;1
1\right\rangle + \frac{\lambda\sqrt{n}\left(\left|n-1;1
0\right\rangle + \left|n-1;0 1\right\rangle\right)}{\omega + E_{0}}
- \frac{\lambda\sqrt{n+1}\left(\left|n+1;1 0\right\rangle +
\left|n+1;0 1\right\rangle\right)}{\omega - E_{0}} \ ,
\end{eqnarray}
%%%%%%%%%%%%%%%%%%%%%%%%%%%%%%%%%%%%%%%%%%%%%%%%%%%%%%%%%%%%%%%%%%%%%%%%%%%%%%%%%%%%%%%%%%%%%%%%%%%%%
where $n$ is the number of photons.

Using Eqs.~(\ref{ham022}) and~(\ref{V22}), we get the eigenvalues of the Hamiltonian $\hat{H}$
in the framework of the second order perturbation theory~\cite{Landau}:
%%%%%%%%%%%%%%%%%%%%%%%%%%%%%%%%%%%%%%%%%%%%%%%%%%%%%%%%%%%%%%%%%%%%%%%%%%%%%%%%%%%%%%%%%%%%%%%%%%%%%
\begin{eqnarray}
\label{ener2} E_{n;0 0}^{(\lambda)} &=& \left(\omega +
\frac{4\lambda^{2}E_{0}}{\omega^{2} - E_{0}^{2}}\right)n -
\frac{2\lambda^{2}}{\omega+ E_{0}} \ , \nonumber \\
E_{n;1 0}^{(\lambda)} &=& E_{n;0 1}^{(\lambda)} = n\omega + E_{0} +
\frac{2 \lambda^{2}\omega} {\omega^{2} - E_{0}^{2}} \ , \nonumber \\
E_{n;1 1}^{(\lambda)} &=& \left(\omega -
\frac{4\lambda^{2}E_{0}}{\omega^{2} - E_{0}^{2}}\right)n + 2 E_{0} -
\frac{2\lambda^{2}}{\omega - E_{0}} \  .
\end{eqnarray}
%%%%%%%%%%%%%%%%%%%%%%%%%%%%%%%%%%%%%%%%%%%%%%%%%%%%%%%%%%%%%%%%%%%%%%%%%%%%%%%%%%%%%%%%%%%%%%%%%%%%%
The second order with respect to qubit-photon coupling $\lambda$
terms in Eqs.~(\ref{ener2}) are not proportional to $n$ and can be
interpreted as Lamb shifts of qubit levels, since the number of photons is not changing due to the static Lamb shift, because the virtual photons are excited and absorbed.
These terms, corresponding to the total Lamb shifts of two qubits in different states are given by the following expressions:
%%%%%%%%%%%%%%%%%%%%%%%%%%%%%%%%%%%%%%%%%%%%%%%%%%%%%%%%%%%%%%%%%%%%%%%%%%%%%%%%%%%%%%%%%%%%%%%%%%%%%
\begin{eqnarray}
\label{Lamb} E_{L,0 0}^{(\lambda)} (\omega) &=& -
\frac{2\lambda^{2}}{\omega+ E_{0}} \ , \nonumber \\
E_{L,1 0}^{(\lambda)}(\omega) &=& E_{L,0 1}^{(\lambda)} =
\frac{2 \lambda^{2}\omega} {\omega^{2} - E_{0}^{2}} \ , \nonumber \\
E_{L,1 1}^{(\lambda)}(\omega) &=& - \frac{2\lambda^{2}}{\omega -
E_{0}} \ .
\end{eqnarray}
%%%%%%%%%%%%%%%%%%%%%%%%%%%%%%%%%%%%%%%%%%%%%%%%%%%%%%%%%%%%%%%%%%%%%%%%%%%%%%%%%%%%%%%%%%%%%%%%%%%%%
In the framework of the model, the total Lamb shift is the sum of the Lamb shifts of the individual qubits.
 The static Lamb shift is obtained by the second order perturbation theory, since it involves the creation and absorption
of the virtual photons by the qubit described by the  Feynman's diagram for the self-energy of the second order. The latter contains two vertexes for the
 qubit-photon coupling~\cite{Bethe}.

If the cavity resonance frequency is changed suddenly from
$\omega_{1}$ to $\omega_{2}$, the cavity photons can be created, and
the qubits can get an excitation due to  change of the parameters of
the cavity.
The amplitudes $A_{n;1 0}^{(L)}$ and  $A_{n;0 1}^{(L)}$ for
excitation of one qubit and creation of  $n$ cavity photons are
given by
%%%%%%%%%%%%%%%%%%%%%%%%%%%%%%%%%%%%%%%%%%%%%%%%%%%%%%%%%%%%%%%%%%%%%%%%%%%%%%%%%%%%%%%%%%%%%%%%%%%%%
\begin{eqnarray}
\label{An10} A_{n;1 0}^{(L)} =  _{\lambda \omega_{2}}\left\langle
n;1 0 \right| \left. 0;0 0 \right\rangle_{\lambda \omega_{1}} \ ,
\hspace{3cm} A_{n;0 1}^{(\lambda)} =  _{\lambda
\omega_{2}}\left\langle n;0 1 \right| \left. 0;0 0
\right\rangle_{\lambda \omega_{1}} \ .
\end{eqnarray}
%%%%%%%%%%%%%%%%%%%%%%%%%%%%%%%%%%%%%%%%%%%%%%%%%%%%%%%%%%%%%%%%%%%%%%%%%%%%%%%%%%%%%%%%%%%%%%%%%%%%%
The amplitude $A_{n;1 1}^{(L)}$ for excitation of two qubits and
creation of  $n$ cavity photons is given by
%%%%%%%%%%%%%%%%%%%%%%%%%%%%%%%%%%%%%%%%%%%%%%%%%%%%%%%%%%%%%%%%%%%%%%%%%%%%%%%%%%%%%%%%%%%%%%%%%%%%%
\begin{eqnarray}
\label{An11} A_{n;1 1}^{(L)} =  _{\lambda \omega_{2}}\left\langle
n;1 1 \right| \left. 0;0 0 \right\rangle_{\lambda \omega_{1}} \ .
\end{eqnarray}
%%%%%%%%%%%%%%%%%%%%%%%%%%%%%%%%%%%%%%%%%%%%%%%%%%%%%%%%%%%%%%%%%%%%%%%%%%%%%%%%%%%%%%%%%%%%%%%%%%%%%

The probabilities of DLE
$w_{n;1 0}^{(L)}$ and $w_{n;0 1}^{(L)}$ of  excitation of one
qubit are given by
%%%%%%%%%%%%%%%%%%%%%%%%%%%%%%%%%%%%%%%%%%%%%%%%%%%%%%%%%%%%%%%%%%%%%%%%%%%%%%%%%%%%%%%%%%%%%%%%%%%%%
\begin{eqnarray}
\label{wn10}  w_{n;1 0}^{(L)} &=& \sum_{n=0}^{\infty} \left|A_{n;1
0}^{(L)}\right|^{2} = \left|A_{1;1 0}^{(L)}\right|^{2} =
\left|_{\lambda \omega_{2}}\left\langle 1;1 0 \right| \left. 0;0 0
\right\rangle_{\lambda \omega_{1}}\right|^{2} =
\lambda^{2}\left(\frac{1}{\omega_{2} + E_{0}} - \frac{1}{\omega_{1}
+ E_{0}}\right)^{2} \nonumber \\ &=&
\lambda^{2}\left(-\frac{E_{L;0
0}^{(\lambda)}(\omega_{2})}{2\lambda^{2}} + \frac{E_{L;0
0}^{(\lambda)}(\omega_{1})}{2\lambda^{2}}\right)^{2} =
\left(\frac{\delta E_{L;0 0}^{(\lambda)}}{2\lambda}\right)^{2} \
, \nonumber \\
w_{n;0 1}^{(L)} &=& \sum_{n=0}^{\infty} \left|A_{n;0
1}^{(L)}\right|^{2} = \left|A_{1;0 1}^{(L)}\right|^{2} =
\left|_{\lambda \omega_{2}}\left\langle 1;0 1 \right| \left. 0;0 0
\right\rangle_{\lambda \omega_{1}}\right|^{2} =
\lambda^{2}\left(\frac{1}{\omega_{2} + E_{0}} - \frac{1}{\omega_{1}
+ E_{0}}\right)^{2} \nonumber \\ &=&
\lambda^{2}\left(-\frac{E_{L;0
0}^{(\lambda)}(\omega_{2})}{2\lambda^{2}} + \frac{E_{L;0
0}^{(\lambda)}(\omega_{1})}{2\lambda^{2}}\right)^{2} =
\left(\frac{\delta E_{L;0 0}^{(\lambda)}}{2\lambda}\right)^{2} \ ,
\end{eqnarray}
%%%%%%%%%%%%%%%%%%%%%%%%%%%%%%%%%%%%%%%%%%%%%%%%%%%%%%%%%%%%%%%%%%%%%%%%%%%%%%%%%%%%%%%%%%%%%%%%%%%%%
where
%%%%%%%%%%%%%%%%%%%%%%%%%%%%%%%%%%%%%%%%%%%%%%%%%%%%%%%%%%%%%%%%%%%%%%%%%%%%%%%%%%%%%%%%%%%%%%%%%%%%%
\begin{eqnarray}
\label{a10} A_{1;1 0}^{(L)} = A_{1;0 1}^{(L)}  =
\lambda\left(\frac{1}{\omega_{2} + E_{0}} - \frac{1}{\omega_{1} +
E_{0}}\right) \ .
\end{eqnarray}
%%%%%%%%%%%%%%%%%%%%%%%%%%%%%%%%%%%%%%%%%%%%%%%%%%%%%%%%%%%%%%%%%%%%%%%%%%%%%%%%%%%%%%%%%%%%%%%%%%%%%
Note that excitation of one qubit is possible only with
creation of one cavity photon.

Let us estimate the probability of excitation of a single qubit due to the DLE by using the physical realization with the superconducting qubits,
coupled to CPW. Following Ref.~\onlinecite{Delsing} we use the parameters: $\omega_{1} = 2\pi \times 5 \ \mathrm{GHz}$,
$E_{0} = 2\pi \times 3.721 \  \mathrm{GHz}$, $\lambda = 0.04 \omega_{1} =  2\pi \times 0.2 \  \mathrm{GHz}$.   According to Ref.~\onlinecite{Johansson},
assuming  the relative change of the Josephson energy and the length of the transmission line, corresponding
to the change of the boundary conditions of the cavity, is $0.25$, we use $\omega_{2} = \omega_{1} - 0.25 \omega_{1}= 2\pi \times 3.75 \  \mathrm{GHz}$.
 Substituting listed above parameters into Eq.~(\ref{wn10}),  one obtains   $ w_{n;1 0}^{(L)} =  w_{n;0 1}^{(L)} = 1.472 \times 10^{-5}$.

The contribution of DLE to the probability $w_{n;1 1}^{(L)}$ of
excitation of two qubits is given by
%%%%%%%%%%%%%%%%%%%%%%%%%%%%%%%%%%%%%%%%%%%%%%%%%%%%%%%%%%%%%%%%%%%%%%%%%%%%%%%%%%%%%%%%%%%%%%%%%%%%%
\begin{eqnarray}
\label{wn11}  w_{n;1 1}^{(L)} = \sum_{n=0}^{\infty} \left|A_{n;1
1}^{(L)}\right|^{2} = \left|A_{0;1 1}^{(L)}\right|^{2} +
\left|A_{2;1 1}^{(L)}\right|^{2} \ ,
\end{eqnarray}
%%%%%%%%%%%%%%%%%%%%%%%%%%%%%%%%%%%%%%%%%%%%%%%%%%%%%%%%%%%%%%%%%%%%%%%%%%%%%%%%%%%%%%%%%%%%%%%%%%%%%
where  $A_{0;1 1}^{(L)}$ is the amplitude of  excitation of two
qubits without creation of cavity photons given by
%%%%%%%%%%%%%%%%%%%%%%%%%%%%%%%%%%%%%%%%%%%%%%%%%%%%%%%%%%%%%%%%%%%%%%%%%%%%%%%%%%%%%%%%%%%%%%%%%%%%%
\begin{eqnarray}
\label{w011} A_{0;1 1}^{(L)} = _{\lambda \omega_{2}}\left\langle 0;1
1 \right| \left. 0;0 0 \right\rangle_{\lambda \omega_{1}} =
  \frac{2\lambda^{2}}{\left(\omega_{1} + E_{0} \right)\left(\omega_{2} - E_{0} \right)}      \ ,
\end{eqnarray}
%%%%%%%%%%%%%%%%%%%%%%%%%%%%%%%%%%%%%%%%%%%%%%%%%%%%%%%%%%%%%%%%%%%%%%%%%%%%%%%%%%%%%%%%%%%%%%%%%%%%%
and  $A_{2;1 1}^{(L)}$ is the amplitude of  excitation of two
qubits with creation of two cavity photons given by
%%%%%%%%%%%%%%%%%%%%%%%%%%%%%%%%%%%%%%%%%%%%%%%%%%%%%%%%%%%%%%%%%%%%%%%%%%%%%%%%%%%%%%%%%%%%%%%%%%%%%
\begin{eqnarray}
\label{w211} A_{2;1 1}^{(L)} = _{\lambda \omega_{2}}\left\langle 2;1
1 \right| \left. 0;0 0 \right\rangle_{\lambda \omega_{1}} =
  -\frac{2\sqrt{2}\lambda^{2}}{\left(\omega_{1} + E_{0} \right)\left(\omega_{2} + E_{0} \right)}      \
  .
\end{eqnarray}
%%%%%%%%%%%%%%%%%%%%%%%%%%%%%%%%%%%%%%%%%%%%%%%%%%%%%%%%%%%%%%%%%%%%%%%%%%%%%%%%%%%%%%%%%%%%%%%%%%%%%
According to Eq.~(\ref{wn11}), the process with excitation of two
qubits is possible only without creation of photons or with creation
of two photons.

Substituting Eqs.~(\ref{w011}) and~(\ref{w211}) into
Eq.~(\ref{wn11}), one gets
%%%%%%%%%%%%%%%%%%%%%%%%%%%%%%%%%%%%%%%%%%%%%%%%%%%%%%%%%%%%%%%%%%%%%%%%%%%%%%%%%%%%%%%%%%%%%%%%%%%%%
\begin{eqnarray}
\label{wnn11}  w_{n;1 1}^{(L)} =
\frac{4\lambda^{4}}{\left(\omega_{1} + E_{0}
\right)^{2}}\left(\frac{1}{\left(\omega_{2} - E_{0}\right)^{2}} +
\frac{2}{\left(\omega_{2} + E_{0}\right)^{2}}\right) =
 \left(E_{L;0 0}^{(\lambda)}(\omega_{1})\right)^{2}\left(\left(\frac{E_{L;1 1}^{(\lambda)}(\omega_{2})}{2\lambda^{2}}\right)^{2}
 + 2 \left(\frac{E_{L;0 0}^{(\lambda)}(\omega_{2})}{2\lambda^{2}}\right)^{2}\right)
 \ .
 \end{eqnarray}
%%%%%%%%%%%%%%%%%%%%%%%%%%%%%%%%%%%%%%%%%%%%%%%%%%%%%%%%%%%%%%%%%%%%%%%%%%%%%%%%%%%%%%%%%%%%%%%%%%%%%
Let us mention that the probabilities of excitation of qubits
are expressed in terms of the Lamb shifts.

 We estimate the probability of excitation of two qubits due to the DLE by using the physical realization with the superconducting qubits,
coupled to CPW. Applying listed above parameters in Eq.~(\ref{wnn11}),  one obtains   $w_{n;1 1}^{(L)} = 0.1$.
According to Eq.~(\ref{wnn11}),   $ w_{n;1 1}^{(L)}$ is enhanced essentially, when $E_{0}$ is near the resonance with $\omega_{2}$.
Let us mention that since the probabilities of excitations of qubits due to the DLE   $w_{n;1 0}^{(L)}$, $w_{n;0 1}^{(L)}$, and $w_{n;1 1}^{(L)}$ are  essentially less than
one, the application of the perturbation theory used above for the system under consideration is valid.

\section{Quantum entanglement for two qubits due to the change of the boundary
conditions in the cavity}

\label{entan}

Note  that the probability of creation of two photons due
to the dynamical Lamb effect in Eq.~(\ref{wnn11}) is not given by
the product of the probabilities of  creation of a single photon
presented in Ref.~\onlinecite{Lozovik}. Therefore, the quantum
entanglement for two qubits occurs due to the dynamical Lamb effect.

It is well-known that the possible measure of quantum entanglement of two
qubits is the concurrence $C$~\cite{Wootters_1997,Wootters_1998,Wootters}, defined for the
two-qubit state $\left|\Phi\right\rangle$
%%%%%%%%%%%%%%%%%%%%%%%%%%%%%%%%%%%%%%%%%%%%%%%%%%%%%%%%%%%%%%%%%%%%%%%%%%%%%%%%%%%%%%%%%%%%%%%%%%%%%
\begin{eqnarray}
\label{Phi} \left|\Phi\right\rangle = a\left|00\right\rangle +
b\left|01\right\rangle + c\left|10\right\rangle +
d\left|11\right\rangle
 \end{eqnarray}
%%%%%%%%%%%%%%%%%%%%%%%%%%%%%%%%%%%%%%%%%%%%%%%%%%%%%%%%%%%%%%%%%%%%%%%%%%%%%%%%%%%%%%%%%%%%%%%%%%%%%
as
%%%%%%%%%%%%%%%%%%%%%%%%%%%%%%%%%%%%%%%%%%%%%%%%%%%%%%%%%%%%%%%%%%%%%%%%%%%%%%%%%%%%%%%%%%%%%%%%%%%%%
\begin{eqnarray}
\label{CPhi} C (\Phi) = 2 \left|ad - bc \right| \ .
 \end{eqnarray}
%%%%%%%%%%%%%%%%%%%%%%%%%%%%%%%%%%%%%%%%%%%%%%%%%%%%%%%%%%%%%%%%%%%%%%%%%%%%%%%%%%%%%%%%%%%%%%%%%%%%%

We define the conditional concurrence for each fixed number of
created photons. For the case of  absence of photons $|0\rangle_{ph}$, we have $a =
A_{0;0 0}^{(L)}$, $d = A_{0;1 1}^{(L)}$,  $b=0$, $c=0$.
For the case of creation one photon $|1\rangle_{ph}$, we have $b =
A_{1;0 1}^{(L)}$, $c = A_{1;1 0}^{(L)}$,  $a=0$, $d=0$. In this
case, the conditional concurrence for one photon
$C_{\left|1\right\rangle}$ is defined as
%%%%%%%%%%%%%%%%%%%%%%%%%%%%%%%%%%%%%%%%%%%%%%%%%%%%%%%%%%%%%%%%%%%%%%%%%%%%%%%%%%%%%%%%%%%%%%%%%%%%%
\begin{eqnarray}
\label{CPhi1} C_{\left|1\right\rangle} = 2 \left|bc \right| = 2
\left|A_{1;0 1}^{(L)} A_{1;1 0}^{(L)}\right| \ .
 \end{eqnarray}
%%%%%%%%%%%%%%%%%%%%%%%%%%%%%%%%%%%%%%%%%%%%%%%%%%%%%%%%%%%%%%%%%%%%%%%%%%%%%%%%%%%%%%%%%%%%%%%%%%%%%
Substituting   Eq.~(\ref{a10})  into Eq.~(\ref{CPhi1}), we obtain
%%%%%%%%%%%%%%%%%%%%%%%%%%%%%%%%%%%%%%%%%%%%%%%%%%%%%%%%%%%%%%%%%%%%%%%%%%%%%%%%%%%%%%%%%%%%%%%%%%%%%
\begin{eqnarray}
\label{CPhi1n} C_{\left|1\right\rangle} = 2\lambda^{2}
\left(\frac{1}{\omega_{2} + E_{0}} - \frac{1}{\omega_{1} +
E_{0}}\right)^{2} = 2 w_{n;1 0}^{(L)} = 2 w_{n;0 1}^{(L)}  = 2
\left(\frac{\delta E_{L;0 0}^{(\lambda)}}{2\lambda}\right)^{2}
  \ .
\end{eqnarray}
%%%%%%%%%%%%%%%%%%%%%%%%%%%%%%%%%%%%%%%%%%%%%%%%%%%%%%%%%%%%%%%%%%%%%%%%%%%%%%%%%%%%%%%%%%%%%%%%%%%%%
 Let us mention that the concurrence is expressed in terms of the
variation of the Lamb shift.

 One can estimate the conditional concurrence for one created photon due to the DLE by using the physical realization with the superconducting qubits,
coupled to CPW. Substituting used above parameters into Eq.~(\ref{CPhi1n}), one obtains $C_{\left|1\right\rangle} = 2.945 \times 10^{-5}$.

For the case of creation two photons $|2\rangle_{ph}$, we have $a =
A_{2;0 0}^{(L)}$, $d = A_{2;1 1}^{(L)}$,  $b=0$, $c=0$. In this
case, the conditional concurrence for two photons
$C_{\left|2\right\rangle}$ is defined as
%%%%%%%%%%%%%%%%%%%%%%%%%%%%%%%%%%%%%%%%%%%%%%%%%%%%%%%%%%%%%%%%%%%%%%%%%%%%%%%%%%%%%%%%%%%%%%%%%%%%%
\begin{eqnarray}
\label{CPhi2} C_{\left|2\right\rangle} = 2 \left|ad \right| = 2
\left|A_{2;0 0}^{(L)} A_{2;1 1}^{(L)} \right| \ .
 \end{eqnarray}
%%%%%%%%%%%%%%%%%%%%%%%%%%%%%%%%%%%%%%%%%%%%%%%%%%%%%%%%%%%%%%%%%%%%%%%%%%%%%%%%%%%%%%%%%%%%%%%%%%%%%
Substituting Eq.~(\ref{w211}) and the amplitude of the creation of  two
photons without excitation of a qubit $A_{2;0 0}^{(L)}$, given by
%%%%%%%%%%%%%%%%%%%%%%%%%%%%%%%%%%%%%%%%%%%%%%%%%%%%%%%%%%%%%%%%%%%%%%%%%%%%%%%%%%%%%%%%%%%%%%%%%%%%%
\begin{eqnarray}
\label{a200} A_{2;0 0}^{(L)} = _{\lambda \omega_{2}}\left\langle 2;0
0 \right| \left. 0;0 0 \right\rangle_{\lambda \omega_{1}} =
  -\frac{2\sqrt{2}\lambda^{2}}{\left(\omega_{1} + E_{0} \right)\left(\omega_{2} - E_{0} \right)}      \
  ,
\end{eqnarray}
%%%%%%%%%%%%%%%%%%%%%%%%%%%%%%%%%%%%%%%%%%%%%%%%%%%%%%%%%%%%%%%%%%%%%%%%%%%%%%%%%%%%%%%%%%%%%%%%%%%%%
 into Eq.~(\ref{CPhi2}), we obtain
%%%%%%%%%%%%%%%%%%%%%%%%%%%%%%%%%%%%%%%%%%%%%%%%%%%%%%%%%%%%%%%%%%%%%%%%%%%%%%%%%%%%%%%%%%%%%%%%%%%%%
\begin{eqnarray}
\label{CPhi2n} C_{\left|2\right\rangle} = \frac{16
\lambda^{4}}{\left(\omega_{1} + E_{0}\right)^{2}\left|\omega_{2}^{2}
- E_{0}^{2} \right|} \ .
\end{eqnarray}
%%%%%%%%%%%%%%%%%%%%%%%%%%%%%%%%%%%%%%%%%%%%%%%%%%%%%%%%%%%%%%%%%%%%%%%%%%%%%%%%%%%%%%%%%%%%%%%%%%%%%

 We estimate the conditional concurrence for two created photons due to the DLE by using the physical realization with the superconducting qubits,
coupled to CPW. Applying previously used parameters in Eq.~(\ref{CPhi2n}),  one obtains   $C_{\left|2\right\rangle} = 1.553\times 10^{-3}$.
According to Eq.~(\ref{CPhi2n}),   $C_{\left|2\right\rangle}$ is enhanced essentially, when $E_{0}$ is near the resonance with $\omega_{2}$. Note that
the probability of excitation of two qubits  and the corresponding conditional concurrence can be essentially enhanced by multiple or periodic
excitation due to change of the boundary conditions. See, e.g.,  Ref.~\onlinecite{Delsing} for another physical realization, namely
 the dynamical Casimir effect for two cavities with variable coupling and
entanglement of qubits, coupled to the cavities; another case is periodic excitation of coupling with cavity~\cite{Lozovik_2015}.
The periodic excitation of the cavity with coupled qubits and also the role of the damping of qubits and cavity will be considered in another study.

Let us mention, that if one of the qubits is in the exited state,
while another one is in the ground state, and the qubits are  in the
state either $\left|1 0 \right\rangle_{qub}$ or $\left|0 1
\right\rangle_{qub}$, there is only one photon in the system. We
conclude that if both qubits are either in the ground state $\left|0
0 \right\rangle_{qub}$ or both qubits are in the excited state
$\left|1 1 \right\rangle_{qub}$, then the photon field in the system
is entangled between the states with absence of photons $\left|0
\right\rangle_{ph}$  and two photons $\left|2 \right\rangle_{ph}$.
The probabilities of each photon state in the cases with both qubits
in the ground $\left|0 0 \right\rangle_{qub}$ and excited $\left|1 1
\right\rangle_{qub}$  states are different. Thus, if both qubits are
in the ground state $\left|0 0 \right\rangle_{qub}$, the probability of two created photons in
the system is given by $\left|A_{2;0 0}^{(L)}\right|^{2}$.  The
amplitude $A_{2;0 0}^{(L)}$ is given by
Eq.~(\ref{a200}). If both qubits are in the exited state $\left|1 1
\right\rangle_{qub}$, the probability of absence of created photons in the
system is given by $\left|A_{0;1 1}^{(L)}\right|^{2}$, and the
probability of two created photons in the system is given by
$\left|A_{2;1 1}^{(L)}\right|^{2}$.  The amplitudes $A_{0;1
1}^{(L)}$ and $A_{2;1 1}^{(L)}$ are given by Eqs.~(\ref{w011})
and~(\ref{w211}), correspondingly.

\section{Discussion and conclusions}

\label{disc}

The superconducting circuits have the advantage of enabling the
study of complex controllable quantum dynamics. This could lead to
quantum simulations and on-chip studies of many-body physics.
 Two-qubit algorithms with a superconducting quantum processor were demonstrated in
Ref.~[\onlinecite{Dicarlo}]. The entangled photon states with the
different frequencies excited due to the DCE in a superconducting
circuit can be applied for the quantum information. The transfer of
signals associated with the photons with the different frequencies
and/or different polarizabilities can be used as the basis for novel
developments in the quantum criptography. These signals can be
created due to the DCE in a superconducting circuit.  We consider
the physical realization to observe the dynamical Lamb effect in the
superconducting qubits which is the novel physical phenomenon, that can be observed experimentally. The numerous novel phenomena and
applications will  be discovered using superconducting
circuits, and these effects will play an important role in future quantum
technologies.

 We evaluated  the quantum
entanglement and the probability of the DLE for two  qubits, coupled to the optical cavity (or the superconducting line for superconducting realizations),
caused by nonadiabatic fast change of the boundary
conditions. We obtained the probability of the dynamical Lamb effect.
According to Eq.~(\ref{wnn11}), the probability of excitation of two
qubits is not equal to the product of the probabilities of
excitation of the first and second qubits. This is the difference
between the dynamic Lamb effect in one and two qubits.
 As  a measure
of the dynamical quantum entanglement, the conditional concurrence for two qubits
for each fixed number of created photons in a nonstationary cavity  was derived and analyzed.
We considered the quantum entanglement not caused by interaction
between two qubits, but due to change of the boundary conditions of
the cavity.

 The system with two qubits is characterized by
their influence on each other and their quantum entanglement with  each other and
created Casimir photons due to the change of the boundary
conditions. This effect is expected to play an important role in
quantum technologies.  The dynamical Lamb effect and quantum entanglement due to
the change of cavity boundary conditions offer a new possibility of  control
of qubits  and open  new directions for quantum technologies.
These phenomena related to the DLE can be observed at the different frequencies regions either for  atoms or for  semiconductor
quantum dots in the optical cavity or
 for the superconducting qubits,  coupled to the superconducting line. The model Hamiltonian, which we use in the present article,  qualitatively describes
the DLE for all three physical realizations.

\acknowledgments

The work was supported by PSC CUNY under Grant No. 66564-00 44.  Yu.~E.~L. was supported by Program of Basic Research of National Research
University HSE.

\end{document}